\documentclass[12pt]{article}  
\usepackage{graphicx}
\usepackage{colordvi}
\usepackage{color}

\newcommand{\be}{\begin{equation}}
\newcommand{\ee}{\end{equation}}
\newcommand{\ea}{\end{eqnarray}}
\newcommand{\ba}{\begin{eqnarray}}

\def\[{\left\lbrack}
\def\]{\right\rbrack}

\def\({\left(}
\def\){\right)}

\begin{document}

%\markboth{G. Oliveira-Neto and L. G. Rezende Rodrigues}
%{Noncommutative cosmological models coupled to phantom fluids}

\title{Noncommutative cosmological models induced by a symplectic formalism coupled to phantom fluids}

\author{G. Oliveira-Neto and L. G. Rezende Rodrigues\\
Departamento de F\'{\i}sica, \\
Instituto de Ci\^{e}ncias Exatas, \\ 
Universidade Federal de Juiz de Fora,\\
CEP 36036-330 - Juiz de Fora, MG, Brazil.\\
gilneto@fisica.ufjf.br, lgrr.fisica.ufjf@gmail.com}

%\author{\footnotesize G. Oliveira-Neto and L. G. Rezende Rodrigues}

%\address{Departamento de F\'{\i}sica, \\
%Instituto de Ci\^{e}ncias Exatas, \\ 
%Universidade Federal de Juiz de Fora,\\
%CEP 36036-330 - Juiz de Fora, MG, Brazil.\\
%gilneto@fisica.ufjf.br, lgrr.fisica.ufjf@gmail.com}
   
\maketitle
%\clearpage
%\pub{Received (Day Month Year)}{Revised (Day Month Year)}

\begin{abstract}
In the present letter, we consider homogeneous and isotropic noncommutative 
cosmological models induced by a symplectic formalism coupled to phantom perfect fluids and a cosmological constant. 
After computing the field equations, we solve them to find the scalar factor dynamics.
We restrict our attention to expansive solutions, that may represent the present
expansion of our Universe. Those solutions generate {\it big rip} singularities. We study
how the parameters, of those models, modify the time it takes to the scalar factor
expands from zero till infinity, at the {\it big rip}.
%\keywords{cosmology; noncommutativity; phantom fluid; accelerated expansion.}
\end{abstract}

%\ccode{PACS Nos.: 04.20.Fy, 04.40.Nr, 11.10.Nx, 98.80.Jk.}

\section{Introduction}
\label{intro}

In 1998 a great discovery took place in cosmology. In a series of observations of distant supernovas two
teams of astronomers discovered that our Universe is undergoing a phase of accelerated expansion \cite{riess,perlmutter}.
If we consider that our Universe is all that exists, that discovery means that some unknow kind of substance, with very 
unusual properties should exists in our Universe. Instead of been attractive, like all other known matter, that substance 
should be repulsive. Besides that, it should be the dominant type of matter present in our Universe. Based on the scale
of the acceleration observed in our Universe, the estimation is that about 70\% of the total matter content of our Universe 
is made of that unknow substance \cite{an}. Since then, many researchers have formulated different proposal for that unknow 
substance, also known in the literature as dark energy. Among them, we may mention some examples: a positive cosmological constant,
phantom fluid, Chaplygin gas, generalized Chaplygin gas, quintessence, quintom, K-essence \cite{bilic,bertolami,caldwell,peebles,caldwell1,copeland,padmanabhan,cai,Mli}.

Many years ago a very interesting idea was introduced in order to try to remove the ultraviolet divergences of quantum fields theories.
That idea was the noncommutativity between the spacetime coordinates, first introduced by H. S. Snyder \cite{snyder}. After the great success
of the renormalization process in quantum field theories, the noncommutative (NC) ideas were forgotten, in that context. Many years later, those
ideas were reborn due to important results in superstring, membrane and $M$-theories. See the following reviews for more information on those 
important results \cite{douglas,schwarz,szabo}. Once those ideas were revived, many physicists took the opportunity to apply them to different
areas like: quantum mechanics, condensed matter, high energy physics and gravity. For a review of those applications see \cite{banerjee}.
Another important area where the noncommutative ideas have been applied is cosmology. In cosmology those ideas were applied to the very early
universe \cite{garcia,nelson,barbosa,sabido,gil} and also the present universe \cite{pedram,obregon,sabido1,sabido2,gil2,gil3,gil1,gil4}. Noncommutativity ideas have also been applied
to black hole physics \cite{sabido3,sabido4,sabido5}. In the very early universe noncommutativity
may have occurred, naturally, due to the very high energies present at that epoch. On the other hand, in the present epoch, it is
possible that some residual noncommutativity may have survived from the very early universe. That residual NC may be the cause of some important 
cosmological physical phenomena, like the present accelerated expansion of our Universe \cite{pedram,obregon,sabido1,sabido2,gil2,gil3,gil1,gil4}.

In the present letter, we study NC homogeneous and isotropic cosmological models coupled to phantom perfect fluids and a cosmological constant.
The spacelike hypersurfaces of the models may have positive ($k=1$), negative ($k=-1$) or nil ($k=0$) constant curvatures.
The noncommutativity is introduced, here, as a deformed algebra of the Poisson brackets between the metric variable, the fluid variable and their conjugate
momenta ($a$, $P_a$, $T$, $P_T$), where $a$ is the scale factor, $T$ is the fluid variable, $P_a$ and $P_T$ are, respectively, their canonically conjugated momenta.
After that, we rewrite the theory in terms of new commutative variables and NC parameters, such that those new variables
satisfy the usual Poisson brackets algebra, up to the
first order in the NC parameters. In the literature, there are two different ways to write the transformations from the NC variables to the new commutative 
variables plus the NC parameters.
In the first one, the transformations are chosen by hand \cite{garcia,nelson,barbosa,gil,pedram,obregon,gil2,gil3}. In the second one, an unique set of transformations is 
induced by the Faddeev-Jackiw (FJ) symplectic formalism \cite{faddeev,barcelos,barcelos1,montani}. That, more elegant, way to introduce the transformations have already been
used in cosmological models \cite{gil0,gil1,gil4}. 
In fact, we want to extend the NC model, introduced in Ref.\cite{gil1}, to the case where the matter content is a phantom perfect fluid.

In the next section, Section \ref{model}, the NC model, we are going to study, is introduced. In Section \ref{results}, we present the main results of the paper. We show
how the noncommutativity introduced, here, modify the commutative model. In Section \ref{estimating}, we estimate the value of the NC parameter. Finally, in Section \ref{conclusions},
we present the main conclusions of the paper.

\section{The general NC model}
\label{model}

Following the authors of Ref.\cite{gil1}, we study, here, NC Friedmann-Robertson-Walker cosmological models. The equation of state of the phantom perfect fluid, present in the model, is given by,
\begin{equation}
\label{1}
p = \alpha \rho,
\end{equation} 
where $p$ is the fluid
pressure, $\rho$ its energy density and $\alpha$ is a constant, such that $\alpha<-1$. The constant curvature of the spatial sections may be positive, negative or zero. The geometrical and matter sectors of the models may be written in their Hamiltonian forms with the aid of the ADM formalism \cite{misner} and the Schutz's variational formalism \cite{schutz,schutz1}, respectively. The authors of Ref.\cite{gil1} used those formalisms and wrote, initially, the commutative Hamiltonian for
a FRW model where the matter content is a perfect fluid with equation of state (\ref{1}) and a cosmological constant ($\Lambda$), for a generic value of $\alpha$. The commutative Hamiltonian is given by,
\begin{equation}
\label{1.3}
H = -\frac{P_{a}^2}{12a} - 3ka +\Lambda a^{3} + P_{T}a^{-3\alpha}\,\,,
\end{equation}
where the authors of Ref.\cite{gil1} used the unit system where $G=c=1$.
In $H$ eq. (\ref{1.3}), $a$ is the scale factor and $P_a$ is its canonically conjugated momentum, they describe the geometrical degrees of freedom of the model.
$T$ is the variable associated to the perfect fluid and $P_T$ is its canonically conjugated momentum, they describe the matter degrees of freedom of the model.
$T$ and $P_T$ are obtained with the aid of the Schutz's variational formalism \cite{schutz,schutz1}. They have already been derived in several papers \cite{rubakov,germano1,gil5,gil2,gil3,gil4}.
The noncommutativity that was studied in Ref.\cite{gil1} and that we are going to study, here,
is not the noncommutativity between spatial coordinates. Rather, it is a noncommutativity
between phase space variables, that are obtained through deformed Poisson brackets between
those variables. Here, the phase space is given by the canonical variables and conjugated
momenta: $\{ a, P_a, T, P_T \}$. Then, the noncommutativity, at the classical level, we 
are going to study will be between these phase space variables. Since these variables 
are functions of the time coordinate $t$, this procedure is a generalization of the
noncommutativity between spatial coordinates.

As we have mentioned before, we want to extend the NC model, introduced in Ref.\cite{gil1}, to the case where the matter content is a phantom perfect fluid. 
In order to obtain a NC version of the FRW model, described by the Hamiltonian (\ref{1.3}), we use the FJ symplectic formalism \cite{faddeev,barcelos,barcelos1,montani}. 
The application of that formalism to the present model is described in details in Ref.\cite{gil1}. Next, let us introduce the most important points of the application of that 
formalism to the present model.

Initially, we have to write the zeroth-iterative Lagrangian of the system,
\begin{equation}
\label{eq:4aa}
{\cal L}^{(0)}={P}_a\dot{a}+P_T\dot{T}-V(a,p_a, T,P_T),
\end{equation}
where $V=N\Omega$ is the symplectic potential, $\Omega$ is the Hamiltonian (\ref{1.3}) and $N$ is a Lagrangian multiplier (also known as the lapse function in the ADM formalism \cite{misner}).
From the Lagrangian (\ref{eq:4aa}), we obtain the symplectic variables: \\$ \xi^{i} = (a, P_a, T, P_T, N), $  
and the corresponding zeroth-iterative one-form canonical momenta,
\begin{equation}
\label{eq:5a}
A_a^{(0)}=P_a,\qquad A_{P_{a}}^{(0)}=0,\qquad A_T^{(0)}=P_T,\qquad
A_{P_{T}}^{(0)}=0,\qquad A_N^{(0)}=0.
\end{equation}
Introducing the symplectic variables $\xi^{i}$ and the one-form canonical momenta (\ref{eq:5a}), in the symplectic matrix definition,
\begin{equation}
\label{eq:7a}
f_{\xi^i\xi^j}=\frac{\partial A_{\xi^j}}{\partial{\xi^i}}-\frac{\partial A_{\xi^i}}{\partial{\xi^j}}.
\end{equation}
we obtain the zeroth-iterative symplectic matrix, given by,
\begin{equation}
\label{smatrix0}
f^{(0)}=\left(\begin{array}{ccccc}
0&-1&0&0&0\\
1&0&0&0&0\\
0&0&0&-1&0\\
0&0&1&0&0\\
0&0&0&0&0\end{array}\right).
\end{equation}
This matrix is singular, which implies the existence of constraints in the model. 
It has the following zero-mode
\begin{equation}
\nu=\left(\begin{array}{ccccc}
0&0&0&0&1
\end{array}\right).
\end{equation}
Multiplying this zero-mode by the symplectic potential gradient we have that, 
\begin{equation}
\label{eq:10a}
\sum_{i=1}^{4}\nu_i\frac{\partial V}{\partial \xi^{i}}=\Omega,
\end{equation}
where $\Omega$ is a constraint. This constraint can be introduced
into the  kinetic  sector of the zeroth-iterative Lagrangian ${\cal L}^{(0)}$, through 
the Lagrangian multiplier $\beta$. In this way, the first-iterative Lagrangian can be written as
\begin{equation}
\label{eq:11a}
{\cal L}^{(1)}={P}_a\dot{a}+P_T\dot{T}+\Omega\dot\beta-N\Omega.
\end{equation}
Now, if we repeat all the steps made in order to find ${\cal L}^{(1)}$ eq. (\ref{eq:11a}) from ${\cal L}^{(0)}$ eq. (\ref{eq:4aa}), we obtain that the first-iterative symplectic matrix is still singular.
It means that the model must have another constraint. When we compute the new constraint, we find that it is identical to the previous one: $\Omega$.
This result indicates that the system has a gauge symmetry that must be fixed and introduced into the zeroth-iterative Lagrangian
(\ref{eq:4aa}), in agreement with the symplectic method \cite{gil0}. In order to fix this symmetry, we introduce a gauge fixing term
$(\Sigma)$ into the original zeroth-iterative Lagrangian (\ref{eq:4aa}), to find the second-iterative Lagrangian,
\begin{eqnarray}
\mathcal{L}^{\left(2\right)} & = & P_{a}\dot{a} + P_{T}\dot{T} + \Sigma\dot{\eta} - N\Omega .
\label{eqL}
\end{eqnarray}
where $\Sigma = N - 1$, which implies that $N = 1$ and $\eta$ is a Lagrange multiplier.
Now, if we repeat all the steps made in order to find the symplectic matrix eq. (\ref{smatrix0}) from ${\cal L}^{(0)}$ eq. (\ref{eq:4aa}), we obtain from $\mathcal{L}^{\left(2\right)}$ eq. (\ref{eqL})
a non-singular second-iterative symplectic matrix. After a straightforward calculation, we obtain the inverse of the second-iterative symplectic matrix: $[f^{(2)}]^{-1}$. It is important to remember that the elements of this matrix corresponds to the Poisson brackets among the symplectic variables: $\xi_{i}^{\left(2\right)} = \left(a, P_{a}, T, P_{T}, N, \eta \right)$.
Now, we are interested in the introduction of the following NC algebra,
\begin{eqnarray}
\label{eq:19a}
{\{}a,T{\}}&=& \theta,\\
\label{eq:19a1}
{\{}P_{a},P_{T}{\}}&=&\beta, \\
\label{eq:19a2}
\left\{a,P_{a}\right\} & = & \left\{T,P_{T}\right\}=1,
\end{eqnarray}
and all the other Poisson brackets are zero. The motivation to introduce those non-trivial Poisson brackets (\ref{eq:19a}) and (\ref{eq:19a1}) is that: since, $a$ and $T$ are the only variables, 
in the configuration space of the theory, it is natural to introduce the NC between them and their conjugated momenta.
We will do that using the matrix $[f^{(2)}]^{-1}$ and carrying out the procedure in a reverse path to obtain the NC space-time Lagrangian.
Therefore, after introducing the nonzero Poisson brackets eqs. (\ref{eq:19a}), (\ref{eq:19a1}) and (\ref{eq:19a2}) in $[f^{(2)}]^{-1}$, we may invert it to obtain the NC symplectic matrix,
\begin{equation}
\label{eq:abc}
f_{NC}= \frac{1}{\beta\theta-1}\left(
\begin{array}{cccccc}
0 & 1 & -\beta & 0 & 0 & 0\\
-1 & 0 & 0 & -\theta &0 & 0\\
\beta & 0 & 0 & 1 & 0 & 0\\
0 & \theta & -1 & 0 & 0 & 0\\
0 & 0 & 0 & 0 & 0 & (\beta\theta-1)\\
0 & 0 & 0 & 0 & (1-\beta\theta) & 0\\
\end{array}
\right) \,\,,
\end{equation}
where $\beta\theta-1\neq 0$. 
Now, we will use the NC symplectic matrix elements (\ref{eq:abc}) and the relations in eq. (\ref{eq:7a}), in order to obtain a set of partial differential equations
for the NC one-form canonical momenta. After solving those partial differential equations, taking in account that the resulting Lagrangian must be free of second
order velocity terms, we find the following NC Lagrangian,
\begin{equation}
\label{eq:27a}
{\cal L} = \frac{1}{1-\beta\theta}(P_{a}-\beta T)\dot{a}  + \frac{1}{1-\beta\theta}P_{T}\dot{T} + \Sigma \dot{\eta} - N\Omega.
\end{equation}
In order to obtain a commutative first-order Lagrangian, we propose the following coordinate transformation in the classical phase space, 
\begin{eqnarray}
\label{eq:29}
\widetilde{P}_{a}=\frac{P_{a}-\beta T}{1-\beta\theta},\qquad \widetilde{P}_{T}=\frac{P_{T}}{1-\beta\theta},\qquad \widetilde{a}=a,\qquad \widetilde{T}=T.\\\nonumber
\end{eqnarray}
The motivation to introduce the new set of variables $\{\tilde{a}, \tilde{T}, \tilde{P_a}, \tilde{P_T}\}$, is that they satisfy, up to the first order in the NC parameters 
($\theta$ and $\beta$) the usual Poisson brackets algebra. When the theory is written in term of those variables, the noncommutativity is described, 
entirely, by the NC parameters. It is important to notice that the transformations leading to the new variables (\ref{eq:29}), were naturally
derived from the Lagrangian (\ref{eq:27a}). It means that the FJ formalism induces, naturally, a set of {\it commuting} variables.
From the transformations, induced by the FJ formalism, for $\widetilde{a}$ and $\widetilde{T}$ eq. (\ref{eq:29}), it is easy to see that,
$\left\{ \tilde{a}, \tilde{T} \right\}  =  \left\{ a, T \right\} = \theta$.
Therefore, the new variables $\tilde{a}$ and $\tilde{T}$ will satisfy the usual Poisson brackets algebra only if $\theta = 0$. Taking that result in account,
we will set $\theta = 0$, which means that the denominator present in eq. (\ref{eq:29}), $1 - \beta \theta$, reduces to 1.
Finally, the modified superhamiltonian of the system is identified as being the symplectic potential $N\Omega$, where $\Omega$ is the Hamiltonian (\ref{1.3}). We may
write that superhamiltonian in terms of the new variables if we invert equations (\ref{eq:29}). Therefore, in the present gauge $N=1$, it leads to,
\begin{equation}
\label{eq:hnc}
H=-\frac{(\widetilde{P}_a+\beta T)^{2}}{12a}-3ka+\Lambda a^3+\widetilde{P}_{T}a^{-3\alpha}.
\end{equation}
Notice that when $\beta=0$ in $H$ eq. (\ref{eq:hnc}), we recover the Hamiltonian eq. (\ref{1.3}).

Computing the Hamilton's equations from the Hamiltonian (\ref{eq:hnc}), we obtain the following equations,
\begin{eqnarray}
\label{eq:dina}
\dot{a} & = & - \frac{1}{6a} \left( \widetilde{P}_{a} + \beta T \right),\\
\label{eq:dinPa} 
\dot{\widetilde{P}}_{a} & = & -\frac{\left( \widetilde{P}_{a} + \beta T\right)^2}{12a^2} + 3k - 3\Lambda a^2 + 3\alpha \widetilde{P}_T a^{-3\alpha -1},\\
\label{eq:dinT}
\dot{T} & = & a^{-3\alpha},\\
\label{eq:dinPT}
\dot{\widetilde{P}}_T & = & \frac{\beta}{6a} \left(\widetilde{P}_{a} + \beta T \right). 
\end{eqnarray}
Now, in order to find the NC Friedmann equation, we start by combining eqs. (\ref{eq:dina}) and (\ref{eq:dinPT}),
\begin{equation}
\dot{\widetilde{P}}_T = -\beta\dot{a}.\\ 
\label{eq:pt}
\end{equation}
Integrating both sides of eq. (\ref{eq:pt}), we find,
\begin{equation}
\widetilde{P}_T = -\beta a + C,\\ 
\label{eq:pt1}
\end{equation}
where $C$ is a positive integration constant which is related to the initial fluid energy density. Finally, introducing eqs. (\ref{eq:dina}) and (\ref{eq:pt1}) in the NC Hamiltonian (\ref{eq:hnc}), and
imposing the constraint equation $H = 0$, for the NC Hamiltonian,
we are able to write the following NC Friedmann equation \cite{gil1},
\begin{equation}
\label{3}
{\dot{a}^2}+V(a)=0,
\end{equation}
where
\begin{equation}
\label{4}
V(a)=k-\frac{1}{3}\Lambda{a^2}+\frac{\beta}{3}a^{-3\alpha} - \frac{C}{3}a^{-3\alpha-1}.
\end{equation}
We may obtain the commutative Friedmann equation by setting $\beta=0$ in $V(a)$ eq. (\ref{4}). 

\section{Results to the NC Friedmann equation}
\label{results}

Now, we shall solve the NC Friedmann equation (\ref{3}) to find the dynamics of the scale factor $a(t)$. We want to concentrate our attention on expansive solutions for they may be
used to explain the present accelerated expansion of our Universe. We, also, want to compare the solutions to the NC equation (\ref{3}) with the solutions to the
corresponding commutative one (obtained from equation (\ref{3}) by setting $\beta=0$ and $C=P_T$). As we can see, $V(a)$ eq. (\ref{4}) depends on the following parameters: 
$k$, which may assume the values +1, 0, -1; $\Lambda$, which may be positive, negative or nil; $C$, which is positive; $\alpha$, which is smaller than -1; and $\beta$, 
which must be small and may be positive, negative or nil. 

Observing $V(a)$ eq. (\ref{4}), we may obtain the qualitative behavior of the solutions to eq. (\ref{3}). We notice that, for $\beta$ positive, the solutions to eq. (\ref{3}) will be bounded. 
It means that, the Universe starts expanding from a small size, then it reaches a maximum size and finally it contracts to a minimum size.
Therefore, in a sense noncommutativity may eliminate the {\it big rip} singularity, when $\beta$ is positive, but it, also, eliminates the expansive solutions.  
Since, we are most interested in expansive solutions, we shall not consider, in the present letter, positive values of $\beta$.
Again, observing $V(a)$ eq. (\ref{4}), this time for models where $\beta$ is negative, we notice, that the solutions to eq. (\ref{3}) will start at a minimum scale factor value, then, they will expand and finally reach an infinity scale factor value at a finite time. Those universes will end in a {\it big rip} singularity. It is important to mention that, the present
noncommutativity, when $\beta$ is negative, does not remove the {\it big rip} singularity, already present in the corresponding
commutative models. 

Now, that we already know the general, qualitative behavior of the solutions to eq. (\ref{3}), when $\beta$ is negative, we want to know the quantitative behavior of those solutions. Therefore, we solved eq. (\ref{3}), numerically, for many different values of all parameters and initial conditions. Studying those solutions, we reach the following conclusions concerning the scale factor behavior: (i) For any value of $\beta$, the scale factor in 
the NC model will always expand more rapidly than in the corresponding commutative one. Also, if we fix all parameters with the exception of $\beta$, the scale factor will expand
more rapidly in the NC model, for smaller values of that parameter; (ii) If we fix all parameters with the exception of $\alpha$, the scale factor will expand
more rapidly in the NC model, as well as, in the corresponding commutative one, for smaller values of that parameter; (iii) If we fix all parameters with the exception of $C$, the 
scale factor will expand more rapidly in the NC model, as well as, in the corresponding commutative one, for bigger values of that parameter; (iv) If we fix all parameters with the 
exception of $\Lambda$, the scale factor will expand more rapidly in the NC model, as well as, in the corresponding commutative one, for bigger values of that parameter; (v) If we 
fix all parameters with the exception of $k$, the scale factor will expand more rapidly in the NC model, as well as, in the corresponding commutative one, for $k=-1$, then for $k=0$
and finally for $k=1$; (vi) The dynamics of the scale factor also depends on the initial value of that quantity ($a(t=0)\equiv a_0$). We notice that, if we fix all parameters and let $a_0$ varies,
the scale factor will expand more rapidly in the NC model, as well as, in the corresponding commutative one, for bigger values of that quantity. One important way to measure the scale 
factor expansion speed is the time it takes to reach the {\it big rip} singularity, from the initial time $t=0$ ($\tau$). Therefore, 
in what follows we shall compute $\tau$ in order to compare the scale factor expansion speed of different models. In the next subsections, we shall present the results for some particular 
models where the above conclusions can be easily verified. The choices of the different parameter and initial values, in those particular models, were made for a better visualization of the results. 
It is important to mention that due to the unit system we are considering, here, the time $\tau$, in the tables of the next subsections, is not given in the usual time units.

\subsection{Model with $k=0$, $\alpha=-1.1$, $\Lambda=0$, $C$ or $P_T=0.5$, $a_0=0.1$}
\label{subsec1}

Let us consider this model in order to exemplify our conclusions concerning the parameter $\beta$.
The choices of the different parameter values, in this particular model, were made for a better visualization of the conclusions presented at page 9.
For this model, the spatial sections are flat and the cosmological constant is nil. The potential $V(a)$ eq. (\ref{4}), for the present model is given by,
\begin{equation}
\label{5}
V(a)=\frac{\beta}{3}a^{3.3} - \frac{1}{6}a^{2.3}.
\end{equation}
This universe starts to expand from a singularity at $a=0$ and then, after a finite time, since $\beta$ is negative, the scale factor goes to infinity. For $\beta=0$, one obtains the 
commutative case. Since, the scale factor exponent in the term with the NC parameter $\beta$ is greater than the one in the phantom perfect fluid term, the time it takes for the scale
factor to reach the {\it big rip} singularity, in the commutative model, is greater than in the NC one. Also, from $V(a)$ eq. (\ref{5}), it is possible to see that for smaller values 
of $\beta$ the scale factor expands more rapidly, in the NC model. Figure 1, shows graphically the scale factor behavior as a function of $t$, for a commutative model 1(a) and five different noncommutative models 1(b). Table \ref{6}, shows $\tau_C$ and $\tau_{NC}$, for each model.

\begin{figure*}
	(a)\includegraphics[width=0.375\textwidth]{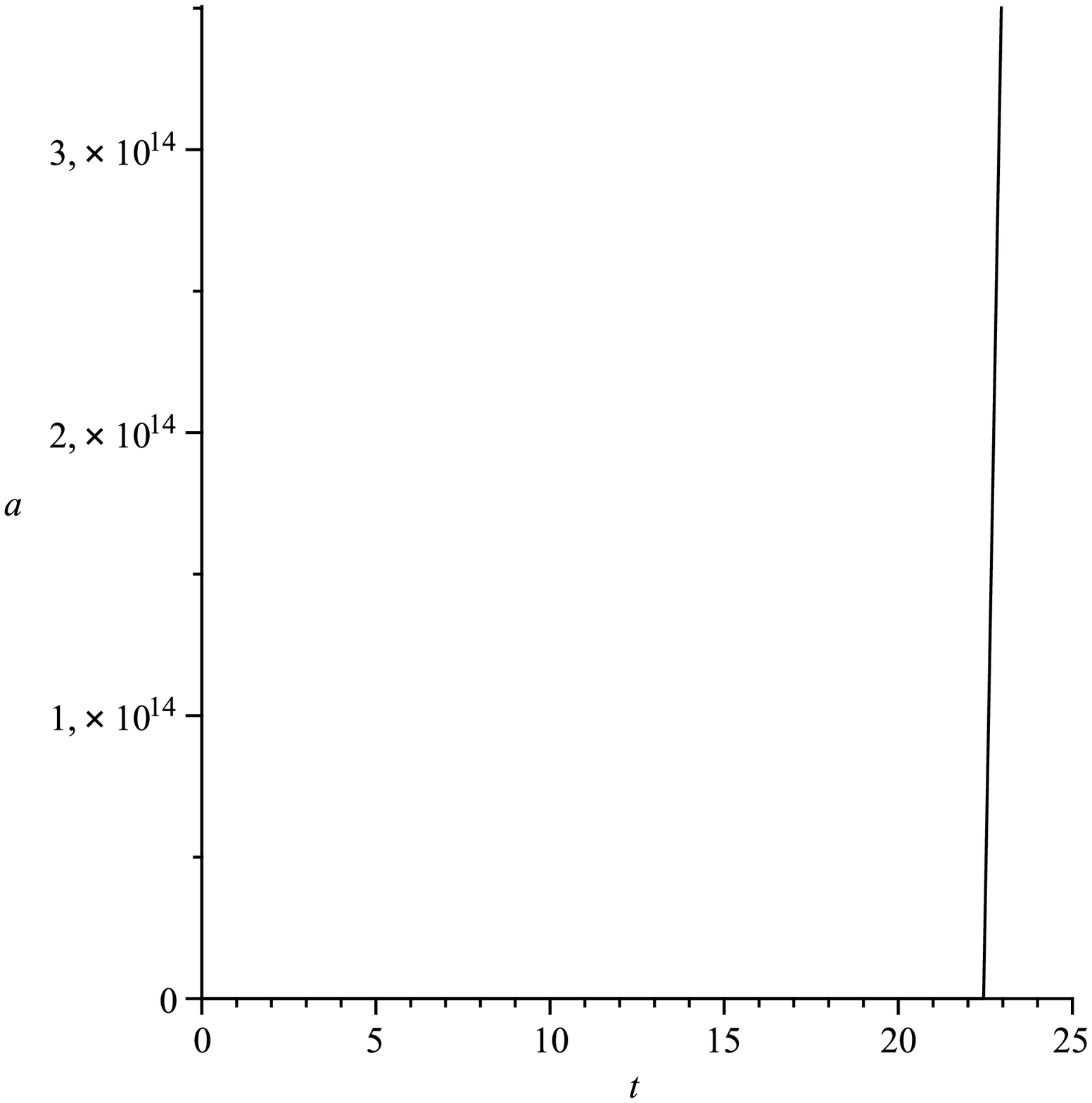}
	(b)\includegraphics[width=0.375\textwidth]{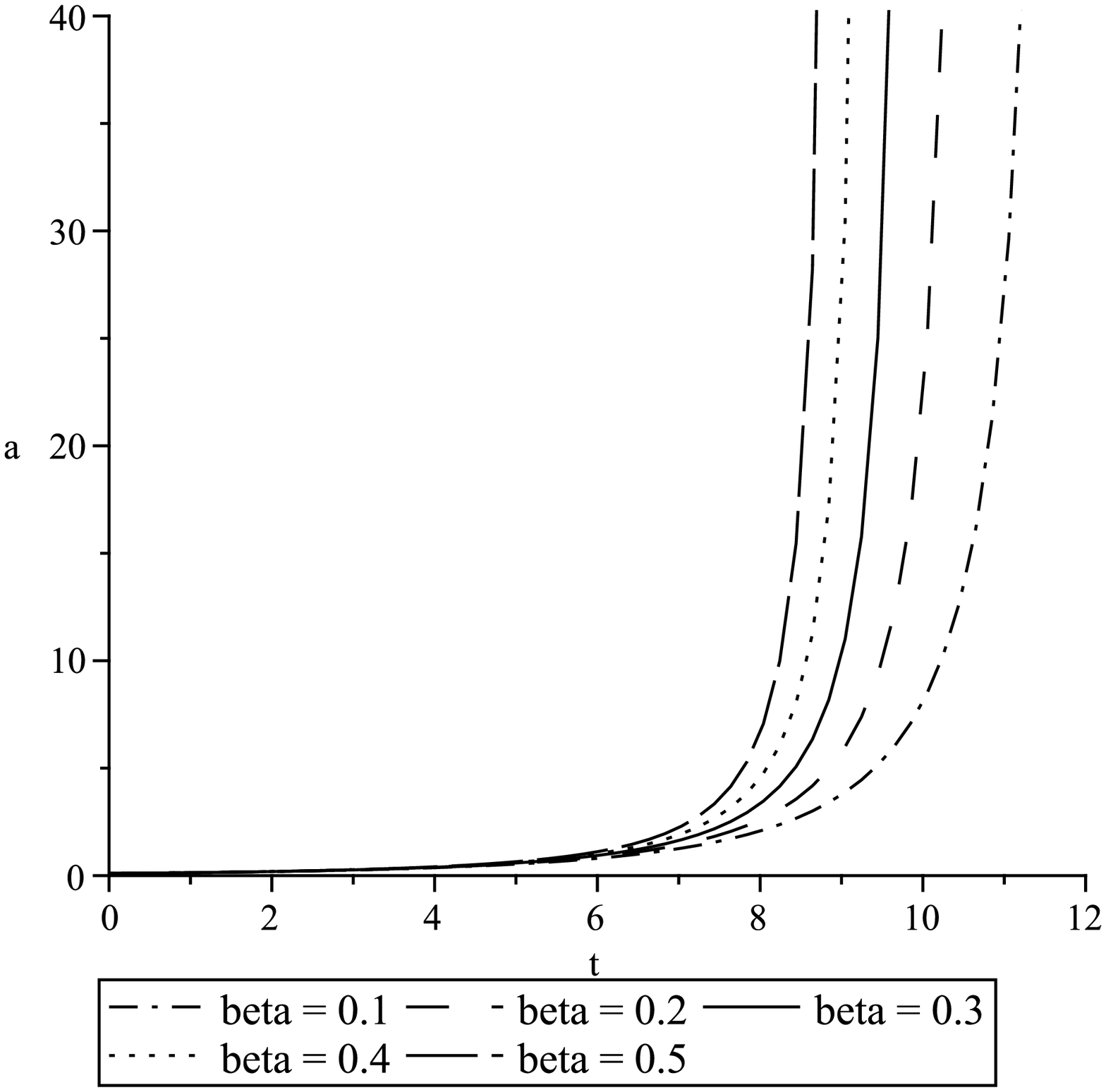}	
	\caption{Comparison between the scale factor dynamics of five examples with different $\beta$ values. 
		{\bf (a)} 
		For the commutative model of Subsection \ref{subsec1}.
		{\bf (b)} 
		For the noncommutative model of Subsection \ref{subsec1}.}
	\label{figure1}
	\end{figure*}

\begin{table}[ht!]
\centering
{\begin{tabular}{|c|c|c|}\hline
			$\beta$ & $\tau_C$ & $\tau_{NC}$ \\ \hline
			-0.1 & $2.3066645\times 10^{1}$ & $1.1957886\times 10^{1}$\\ \hline
			-0.2 & $2.3066645\times 10^{1}$ & $1.0776161\times 10^{1}$\\ \hline
			-0.3 & $2.3066645\times 10^{1}$ & $1.0039779\times 10^{1}$\\ \hline
			-0.4 & $2.3066645\times 10^{1}$ & 9.4989056\\ \hline
			-0.5 & $2.3066645\times 10^{1}$ & 9.0700437\\ \hline
\end{tabular}}
\caption{{\protect\footnotesize The table shows $\tau_{NC}$ and $\tau_C$ for five examples, with different $\beta$ values, of the model \ref{subsec1}.}}
\label{6}
\end{table}

\subsection{Model with $k=0$, $\beta=-0.3$, $\Lambda=-5$, $C$ or $P_T=3$, $a_0=1.5$}
\label{subsec2}

Let us consider this model in order to exemplify our conclusions concerning the parameter $\alpha$.
The choices of the different parameter values, in this particular model, were made for a better visualization of the conclusions presented at page 9.
For this model, the spatial sections are flat and the cosmological constant is negative. The potential $V(a)$ eq. (\ref{4}), for the present model is given by,
\begin{equation}
\label{7}
V(a)=\frac{5}{3}{a^2} - \frac{1}{10}a^{-3\alpha} - a^{-3\alpha-1}.
\end{equation}
Due to the $\Lambda$ term, which gives a positive contribution to the potential $V(a)$ eq. (\ref{7}), $V(a)$ develops a barrier that starts at $a=0$ and ends at the unique potential root ($a_r>0$).
It means that, the universe starts to expand, in that model, from a scale factor value that is greater than or, at least, equal to $a_r$. Then, after a finite time the scale factor goes to infinity. 
Now, if we consider models with different values of $\alpha$, we may compute $\tau$ for those models and verify that for models with smaller values of $\alpha$, the scale factor expands more rapidly. 
For the NC case, as well as for the corresponding commutative one. Figure 2, shows graphically the scale factor behavior as a function of $t$, for five different commutative 2(a) and noncommutative 2(b) models. Table \ref{8}, shows $\tau_C$ and $\tau_{NC}$, for each model.

\begin{figure*}
	(a)\includegraphics[width=0.375\textwidth]{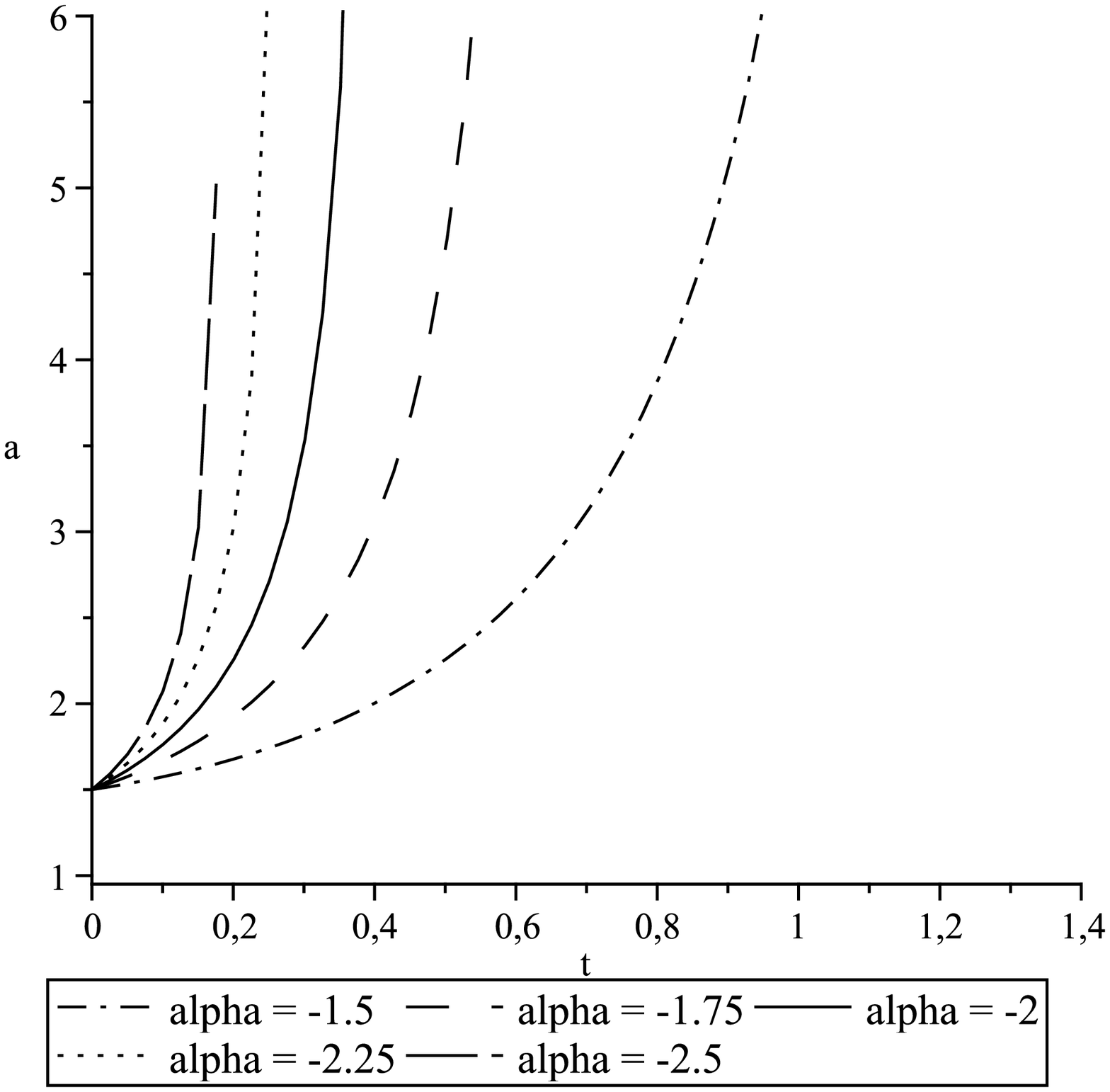}
	(b)\includegraphics[width=0.375\textwidth]{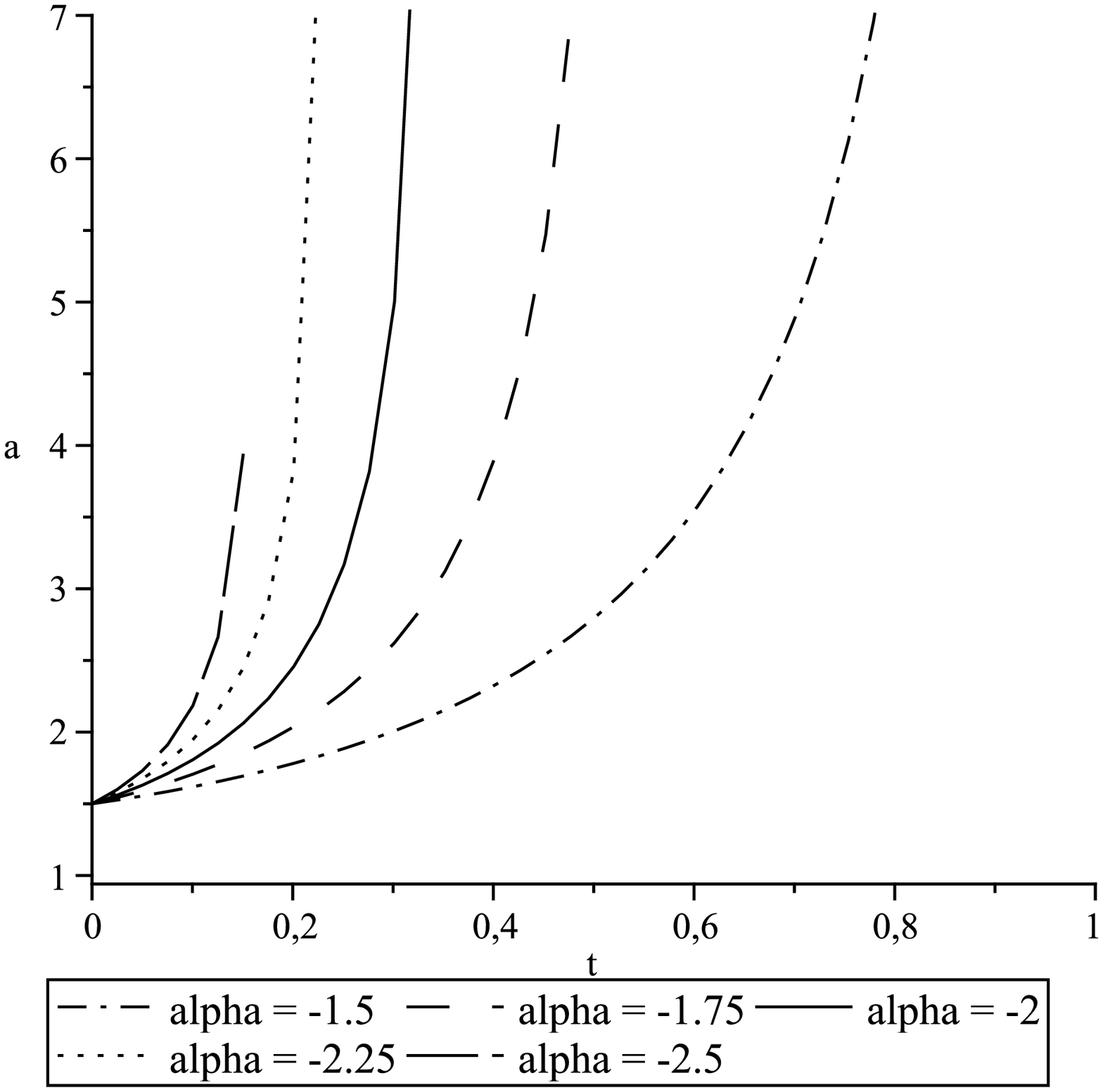}	
	\caption{Comparison between the scale factor dynamics of five examples with different $\alpha$ values. 
		{\bf (a)} 
		For the commutative model of Subsection \ref{subsec2}.
		{\bf (b)} 
		For the noncommutative model of Subsection \ref{subsec2}.}
	\label{figure2}
	\end{figure*}

\begin{table}[ht!]
\centering
\begin{tabular}{|c|c|c|}\hline
			$\alpha$ & $\tau_C$ & $\tau_{NC}$ \\ \hline
			-1.5 & 1.3026421 & 0.95089510\\ \hline
			-1.75 & 0.65971531 & 0.53859025\\ \hline
			-2.0 & 0.40239020 & 0.34367680\\ \hline
			-2.25 & 0.26770709 & 0.23425704\\ \hline
			-2.5 & 0.18764312 & 0.16663763\\ \hline
		\end{tabular}
    \caption{{\protect\footnotesize The table shows $\tau_{NC}$ and $\tau_C$ for five examples, with different $\alpha$ values, of the model \ref{subsec2}.}}
    \label{8}
\end{table}

\subsection{Model with $k=1$, $\alpha=-1.2$, $\beta=-0.1$, $\Lambda=3$, $a_0=0.9$}
\label{subsec3}

Let us consider this model in order to exemplify our conclusions concerning the parameters $P_T$ (commutative) or $C$ (noncommutative).
The choices of the different parameter values, in this particular model, were made for a better visualization of the conclusions presented at page 9.
For this model, the spatial sections have constant positive curvatures and the cosmological constant is positive. The potential $V(a)$ eq. (\ref{4}), for the present model is given by,
\begin{equation}
\label{9}
V(a)=1 - {a^2} - \frac{1}{30}a^{3.6} - \frac{C}{3}a^{2.6}.
\end{equation}
In the present model, at $a=0$, the potential $V(a)$ eq. (\ref{9}) is positive and equal to one, due to the constant positive curvature of the spatial sections. $V(a)$ has an unique positive root ($a_r$).
It means that, the universe starts to expand, in that model, from a scale factor value that is greater than or, at least, equal to $a_r$. Then, after a finite time the scale factor goes to infinity. 
Now, if we consider models with different values of $P_T$ or $C$, we may compute $\tau$ for those models and verify that for models with greater values of $P_T$ or $C$, the scale factor expands more rapidly. For the NC 
case, as well as for the corresponding commutative one. Figure 3, shows graphically the scale factor behavior as a function of $t$, for five different commutative 3(a) and noncommutative 3(b) models. Table \ref{10}, shows $\tau_C$ and $\tau_{NC}$, for each model.

\begin{figure*}
	(a)\includegraphics[width=0.375\textwidth]{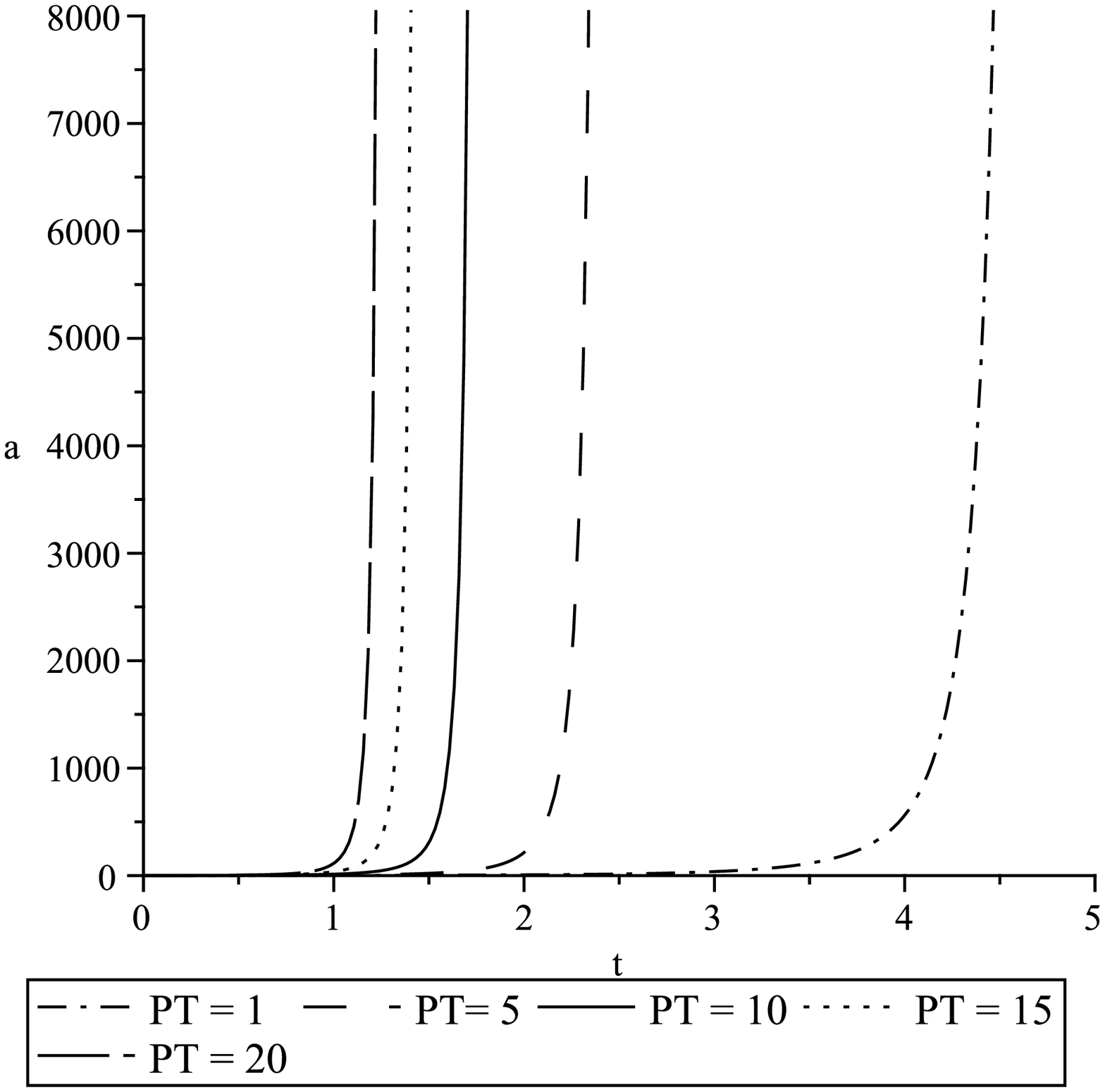}
	(b)\includegraphics[width=0.375\textwidth]{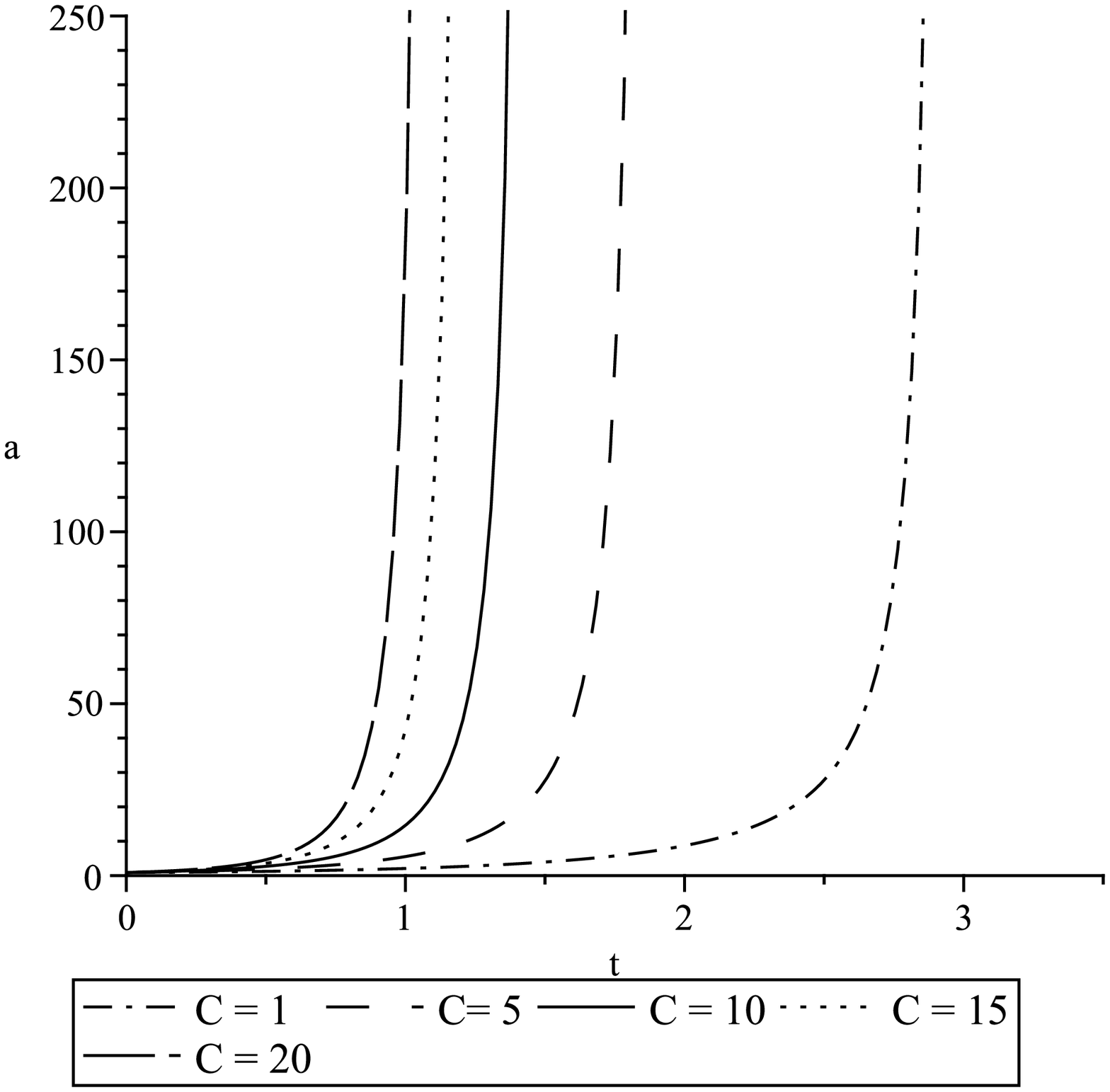}	
	\caption{Comparison between the scale factor dynamics of five examples with different $P_T$ or $C$ values. 
		{\bf (a)} 
		For the commutative model of Subsection \ref{subsec3}.
		{\bf (b)} 
		For the noncommutative model of Subsection \ref{subsec3}.}
	\label{figure3}
	\end{figure*}

\begin{table}[ht!]
\centering
\begin{tabular}{|c|c|c|c|c|}\hline
			 $C$ or $P_T$ & $\tau_C$ & $\tau_{NC}$ \\ \hline
			 1 & 4.8549648 & 2.9386246\\ \hline
			 5 & 2.5141775 & 1.8683459\\ \hline
			 10 & 1.8271527 & 1.4456577\\ \hline
			 15 & 1.5066382 & 1.2283837\\ \hline
			 20 & 1.3114317 & 1.0894517\\ \hline
		\end{tabular}
    \caption{{\protect\footnotesize The table shows $\tau_{NC}$ and $\tau_C$ for five examples, with different $P_T$ or $C$ values, of the model \ref{subsec3}.}}
    \label{10}
\end{table}

\subsection{Model with $k=1$, $\alpha=-1.5$, $\beta=-0.5$, $C$ or $P_T=1$, $a_0=4$}
\label{subsec4}

Let us consider this model in order to exemplify our conclusions concerning the parameter $\Lambda$.
The choices of the different parameter values, in this particular model, were made for a better visualization of the conclusions presented at page 9.
For this model, the spatial sections have constant positive curvatures. The potential $V(a)$ eq. (\ref{4}), for the present model is given by,
\begin{equation}
\label{11}
V(a)=1 - \frac{1}{3}\Lambda{a^2} - \frac{1}{6}a^{4.5} - \frac{1}{3}a^{3.5}.
\end{equation}
In the present model, at $a=0$, the potential $V(a)$ eq. (\ref{11}) is positive and equal to one, as in the previous case, due to the constant positive curvature of the spatial sections. 
$V(a)$ has an unique positive root ($a_r$). It means that, the universe starts to expand, in that model, from a scale factor value that is greater than or, at least, equal to $a_r$. Then, after a finite time the scale factor goes to infinity. Now, if we consider models with different values of $\Lambda$, we may compute $\tau$ for those models and verify that for models with greater values of $\Lambda$, the scale factor expands more rapidly. For the NC case, as well as for the corresponding commutative one. That result is valid for positive as well as for negative values of that parameter. 
Figure 4, shows graphically the scale factor behavior as a function of $t$, for six different commutative 4(a) and noncommutative 4(b) models. Table \ref{12}, shows $\tau_C$ and $\tau_{NC}$, for each model.

\begin{figure*}
	(a)\includegraphics[width=0.375\textwidth]{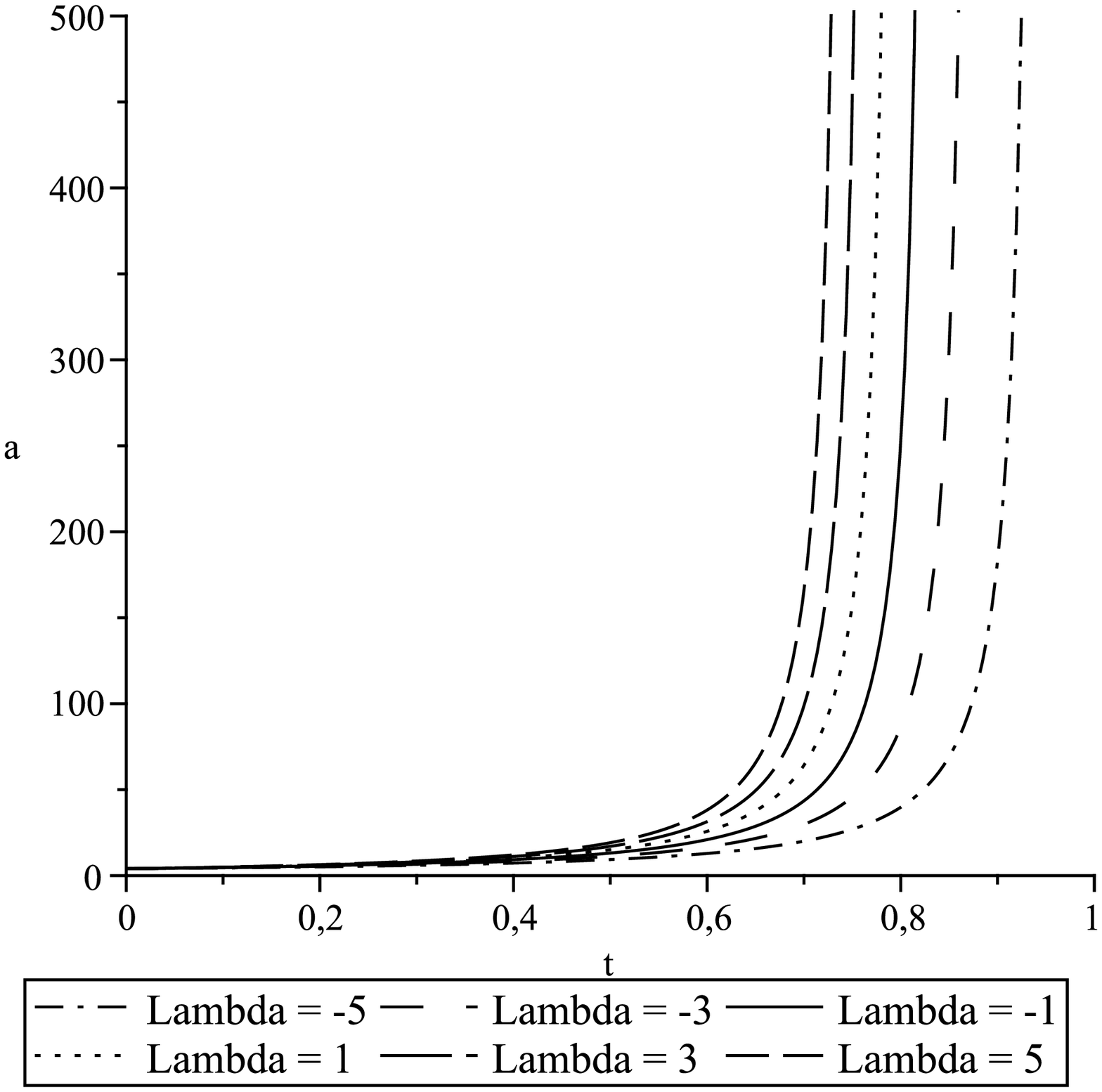}
	(b)\includegraphics[width=0.375\textwidth]{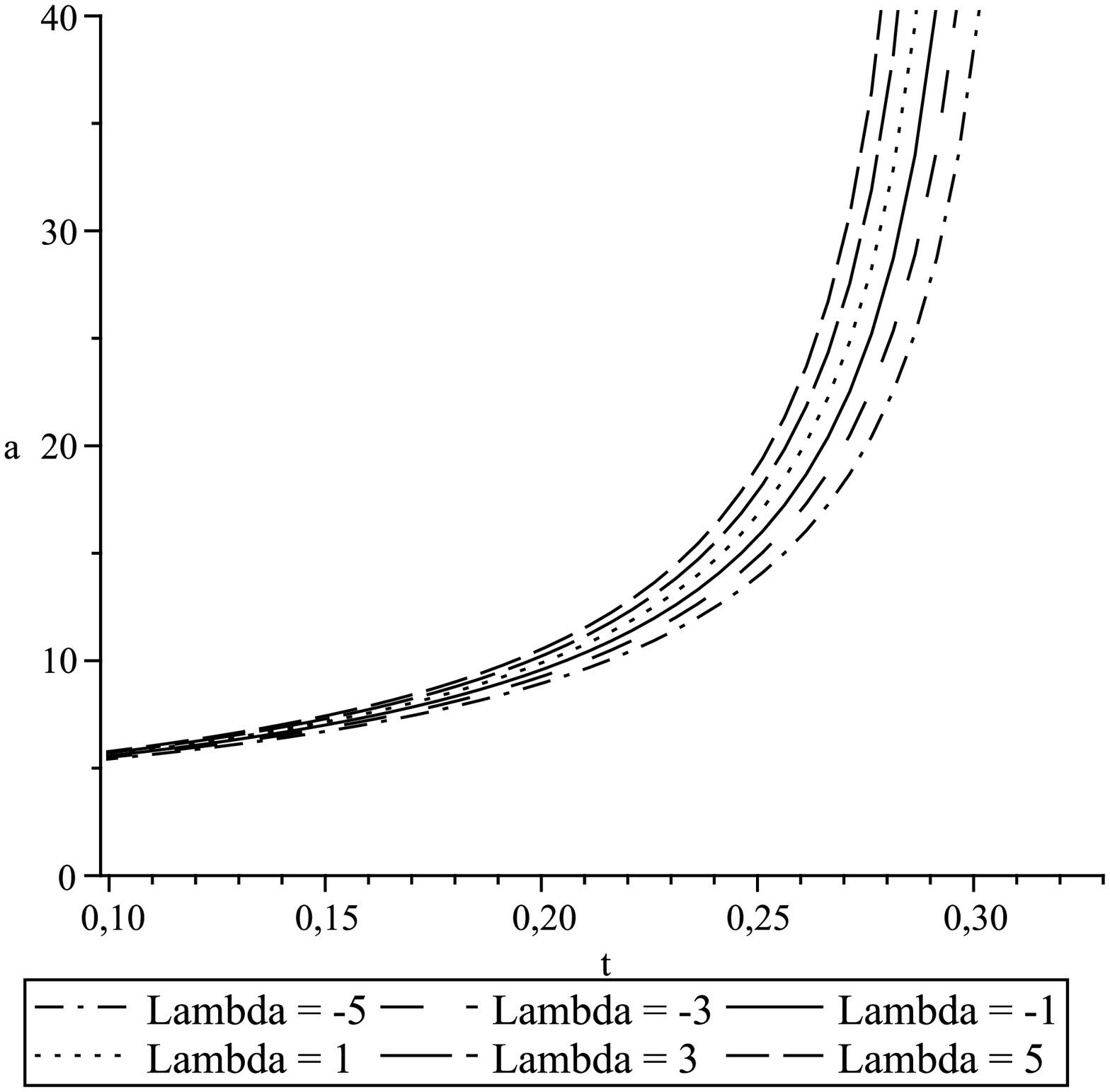}	
	\caption{Comparison between the scale factor dynamics of six examples with different $\Lambda$ values. 
		{\bf (a)} 
		For the commutative model of Subsection \ref{subsec4}.
		{\bf (b)} 
		For the noncommutative model of Subsection \ref{subsec4}.}
	\label{figure4}
	\end{figure*}

\begin{table}[ht!]
\centering
\begin{tabular}{|c|c|c|c|c|}\hline
			 $\Lambda$ & $\tau_C$ & $\tau_{NC}$ \\ \hline
			 -5 & $9.4644114\times 10^{-1}$ & $3.2040206\times 10^{-1}$\\ \hline
			 -3 & $8.8158718\times 10^{-1}$ & $3.1516455\times 10^{-1}$\\ \hline
			 -1 & $8.3651902\times 10^{-1}$ & $3.1035841\times 10^{-1}$\\ \hline
			  1 & $8.0187070\times 10^{-1}$ & $3.0591836\times 10^{-1}$\\ \hline
			  3 & $7.7372657\times 10^{-1}$ & $3.0179310\times 10^{-1}$\\ \hline
			  5 & $7.5004911\times 10^{-1}$ & $2.9794151\times 10^{-1}$\\ \hline
		\end{tabular}
    \caption{{\protect\footnotesize The table shows $\tau_{NC}$ and $\tau_C$ for six examples, with different $\Lambda$ values, of the model \ref{subsec4}.}}
    \label{12}
\end{table}

\subsection{Model with $\Lambda=0$, $\alpha=-2$, $\beta=-0.2$, $C$ or $P_T=0.1$, $a_0=2$}
\label{subsec5}

Let us consider this model in order to exemplify our conclusions concerning the parameter $k$.
The choices of the different parameter values, in this particular model, were made for a better visualization of the conclusions presented at page 9.
For this model, the constant curvatures of the spatial sections may be positive, negative or nil.
The potential $V(a)$ eq. (\ref{4}), for the present model is given by,
\begin{equation}
\label{13}
V(a)=k - \frac{1}{15}a^{6} - \frac{1}{30}a^{5}.
\end{equation}
In the present model, at $a=0$, the potential $V(a)$ eq. (\ref{13}) may have the value one, minus one or nil. That will depend on the value of the constant curvature of the spatial sections. 
$V(a)$ may have no roots (for $k=-1$), one root at $a_r=0$ (for $k=0$) or one root at $a_r>0$ (for $k=1$). It means that, the universe starts to expand, in that model, from $a=0$ (for $k=-1$ or $k=0$)
or from a scale factor value that is greater than or, at least, equal to $a_r>0$ (for $k=1$). Then, after a finite time the scale factor goes to infinity. Now, if we consider models with different values 
of $k$, we may compute $\tau$ for those models and verify that the scale factor expands more rapidly for $k=-1$, then for $k=0$ and finally for $k=1$. For the NC case, as well as for the corresponding
commutative one. Figure 5, shows graphically the scale factor behavior as a function of $t$, for three different commutative 5(a) and noncommutative 5(b) models. Table \ref{14}, shows $\tau_C$ and $\tau_{NC}$, for each model.

\begin{figure*}
	(a)\includegraphics[width=0.375\textwidth]{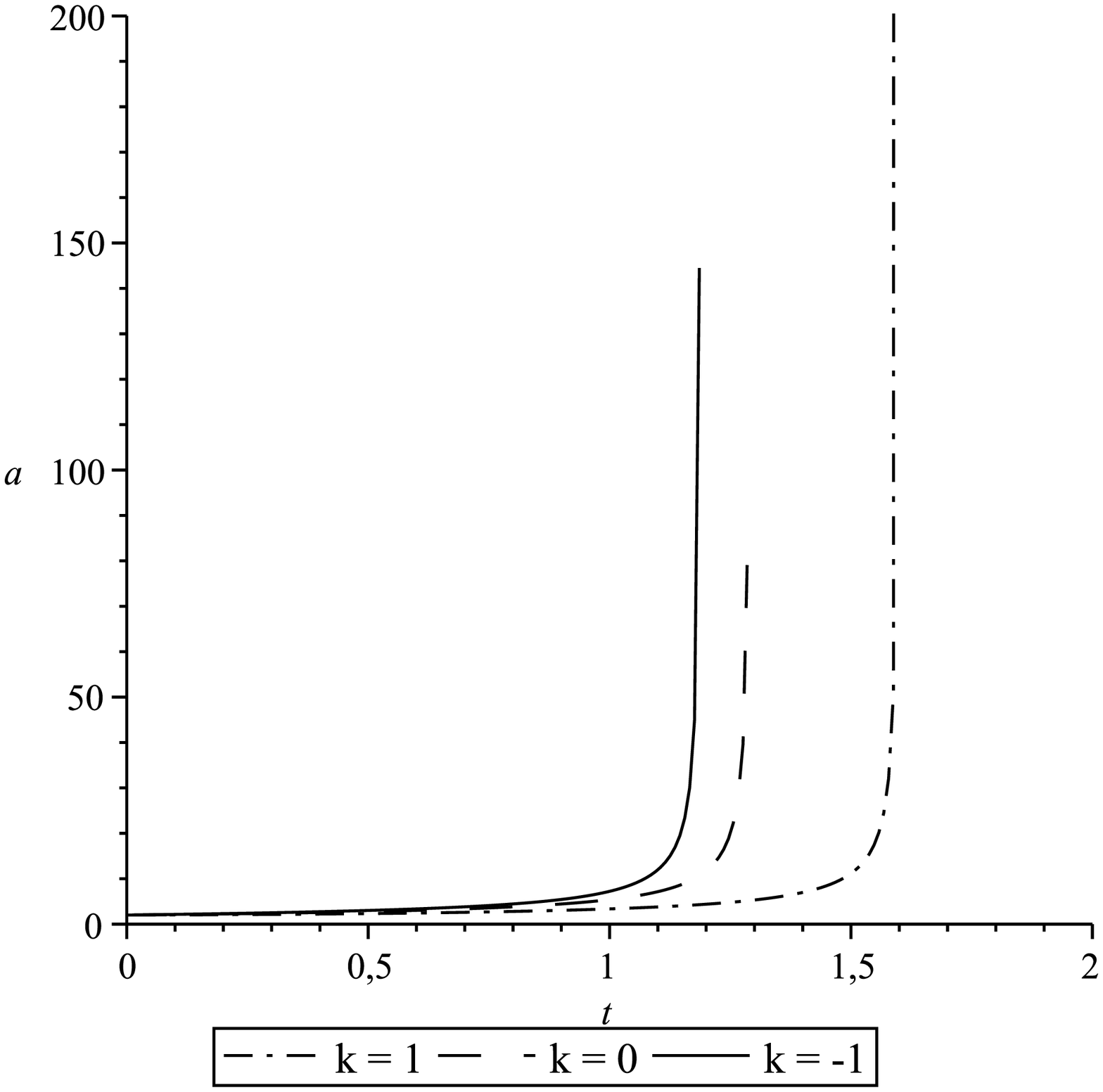}
	(b)\includegraphics[width=0.375\textwidth]{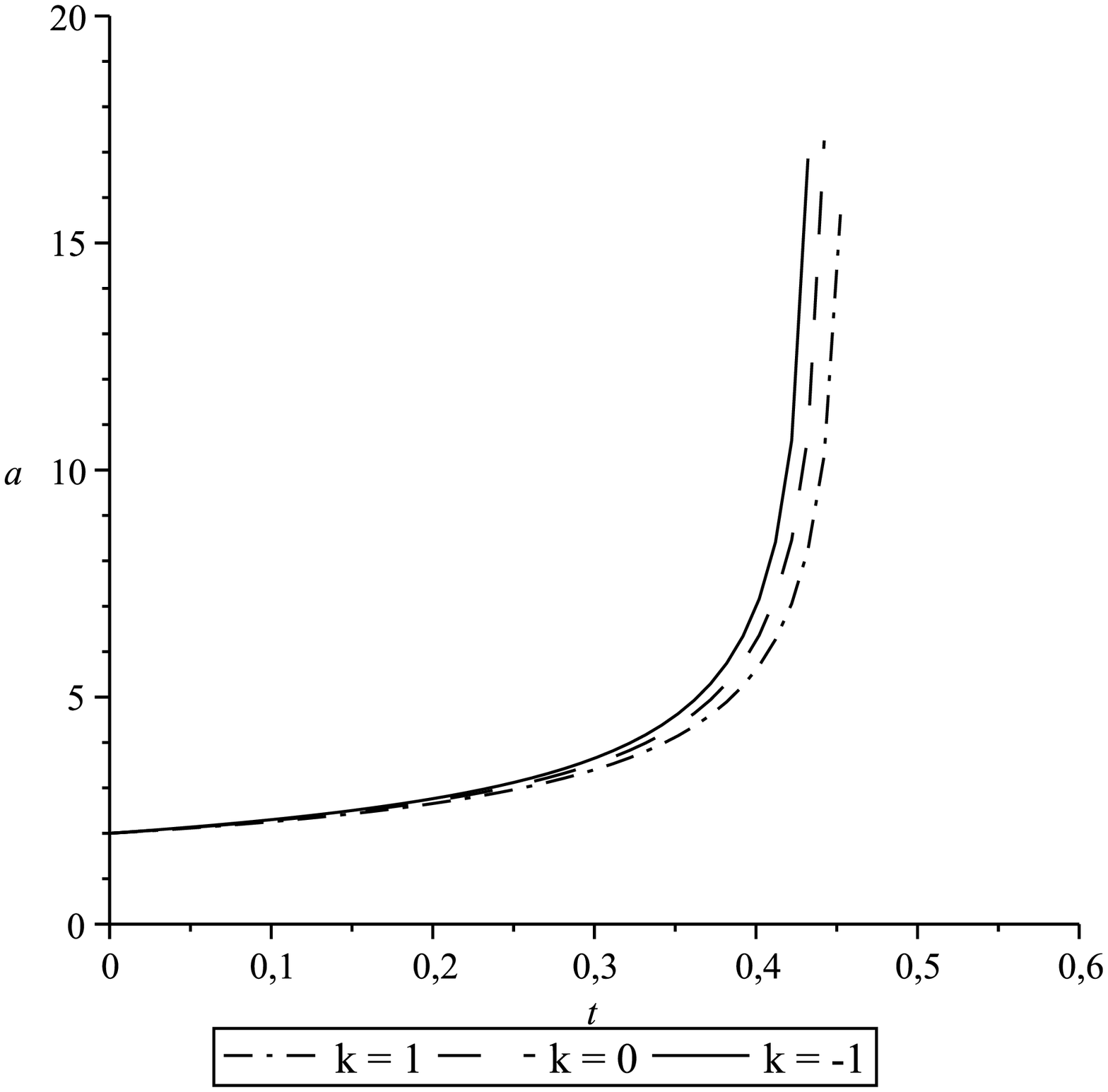}	
	\caption{Comparison between the scale factor dynamics of three examples with different $k$ values. 
		{\bf (a)} 
		For the commutative model of Subsection \ref{subsec5}.
		{\bf (b)} 
		For the noncommutative model of Subsection \ref{subsec5}.}
	\label{figure5}
	\end{figure*}

\begin{table}[ht!]
\centering
\begin{tabular}{|c|c|c|c|c|}\hline
			 $k$ & $\tau_C$ & $\tau_{NC}$ \\ \hline
			 -1 & 1.1880318 & $4.3890845\times 10^{-1}$\\ \hline
			  0 & 1.2909943 & $4.4865168\times 10^{-1}$\\ \hline
			  1 & 1.5980050 & $4.6010663\times 10^{-1}$\\ \hline
		\end{tabular}
    \caption{{\protect\footnotesize The table shows $\tau_{NC}$ and $\tau_C$ for three examples, with different $k$ values, of the model \ref{subsec5}.}}
    \label{14}
\end{table}

\subsection{Model with $k=-1$, $\Lambda=-5$, $\alpha=-1.5$, $\beta=-0.5$, $C$ or $P_T=1$}
\label{subsec6}

Let us consider this model in order to exemplify our conclusions concerning the initial condition $a_0$.
The choices of the different parameter values, in this particular model, were made for a better visualization of the conclusions presented at page 9.
For this model, the spatial sections have constant negative curvatures.
The potential $V(a)$ eq. (\ref{4}), for the present model is given by,
\begin{equation}
\label{15}
V(a)= - 1 + \frac{5}{3}{a^2} - \frac{1}{6}a^{4.5} - \frac{1}{3}a^{3.5}.
\end{equation}
Due to the $\Lambda$ term, which gives a positive contribution to the potential $V(a)$ eq. (\ref{15}), $V(a)$ develops a barrier that starts at its first positive root $a_{r1}$ and ends at its second
positive root $a_{r2}$.
Where $a_{r2} > a_{r1}$. Therefore, if the universe starts to expand with a initial scale factor $a_0 < a_{r1}$, it will reach a maximum value and then it will contract to a {\it big crunch} singularity,
at $a=0$. Since, we are only interested in expansive solutions, in the present letter, we must consider models where the universe starts to expand with a initial scale factor $a_0 \geq a_{r2}$. If it is true,
after a finite time the scale factor goes to infinity. Now, if we consider models with different values of $a_0$, we may compute $\tau$ for those models and verify that for models with greater values of
$a_0$, the scale factor expands more rapidly. For the NC case, as well as for the corresponding commutative one. Figure 6, shows graphically the scale factor behavior as a function of $t$, for five different commutative 6(a) and noncommutative 6(b) models. Table \ref{16}, shows $\tau_C$ and $\tau_{NC}$, for each model.

\begin{figure*}
	(a)\includegraphics[width=0.375\textwidth]{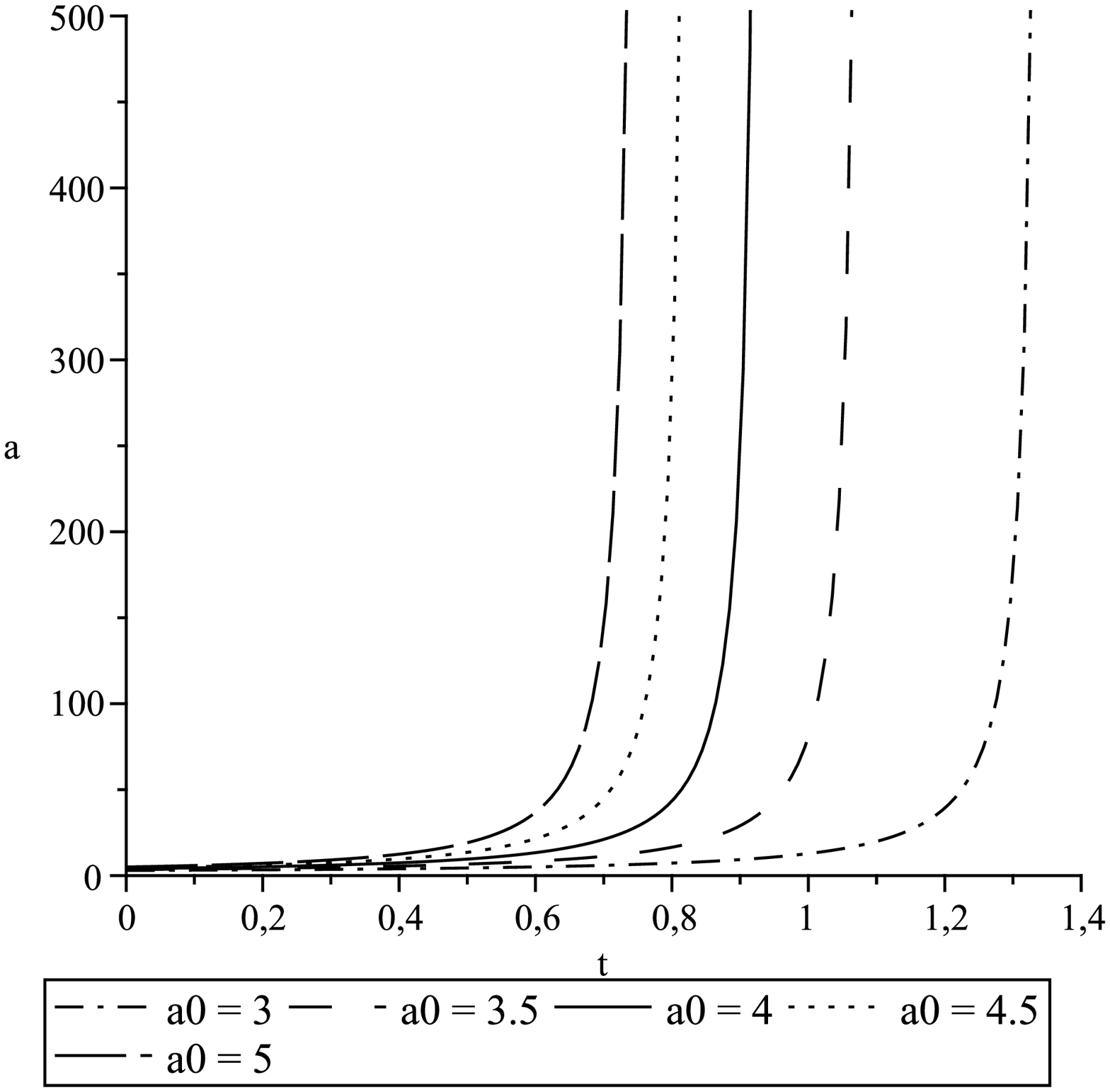}
	(b)\includegraphics[width=0.375\textwidth]{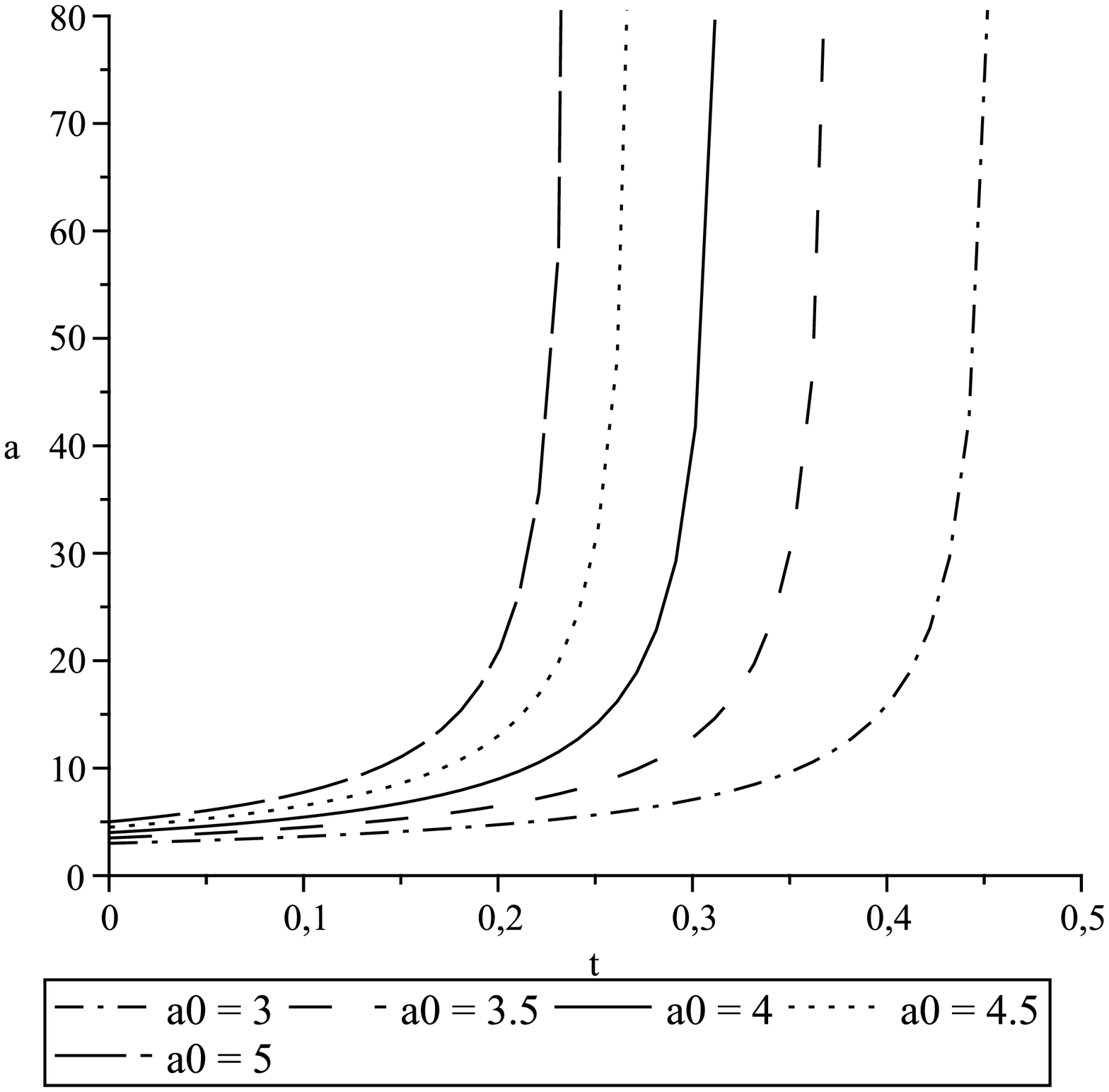}	
	\caption{Comparison between the scale factor dynamics of five examples with different $a0$ values. 
		{\bf (a)} 
		For the commutative model of Subsection \ref{subsec6}.
		{\bf (b)} 
		For the noncommutative model of Subsection \ref{subsec6}.}
	\label{figure6}
	\end{figure*}

\begin{table}[ht!]
\centering
\begin{tabular}{|c|c|c|c|c|}\hline
			 $a_0$ & $\tau_C$ & $\tau_{NC}$ \\ \hline
			 3.0 & 1.3477826 & $4.6008098\times 10^{-1}$\\ \hline
			 3.5 & 1.0859040 & $3.7768274\times 10^{-1}$\\ \hline
			 4.0 & $9.3707133\times 10^{-1}$ & $3.1973733\times 10^{-1}$\\ \hline
			 4.5 & $8.3351560\times 10^{-1}$ & $2.7662077\times 10^{-1}$\\ \hline
			 5.0 & $7.5531487\times 10^{-1}$ & $2.4325411\times 10^{-1}$\\ \hline
		\end{tabular}
    \caption{{\protect\footnotesize The table shows $\tau_{NC}$ and $\tau_C$ for five examples, with different $a_0$ values, of the model \ref{subsec6}.}}
    \label{16}
\end{table}

\section{Estimating the value of the noncommutative parameter $\beta$}
\label{estimating}

In the present section, we want to estimate the value of the NC parameter $\beta$. In order to do that let us consider
the NC Friedmann equation (\ref{3}) with the parameters fixed, by the present observations. We, must also, consider the 
constants G and c in their MKS units, since we want to obtain the time $\tau$ in usual time units. Initially, we fix $k=0$, which
is the most accepted value for the curvature parameter, currently. Then, we consider that the only dark energy source is
the phantom perfect fluid. Therefore, we set $\Lambda=0$. We fix the perfect fluid parameter $\alpha=-1.01$, which agrees
with the observations \cite{riess1}. The parameter $C$ may be written as $C=\Omega_{de}H_0^2$, where $\Omega_{de}$ is the
dark energy mass parameter and $H_0$ is the present Hubble constant. We use, here, that $\Omega_{de}=0.7$ and 
$H_0=72kms^{-1}Mpc^{-1}$ \cite{riess1}. Introducing those values for the parameters in eq. (\ref{3}), we may write it as
an integral equation, after separating the scale factor dependence from the time dependence,
\begin{eqnarray}
\label{17}
\int_{a_{h}}^{1} \frac{da}{\sqrt{ \left( (4.614\times 10^{-36})a^{2.02} - 3 \beta a^{3.03}\right)}} & = & \frac{1}{3} \int_{t_{h}}^{4.32\times 10^{17}} dt.
\end{eqnarray}

In eq. (\ref{17}), we fix the present scale factor value $a_0=1$ and the present age of the Universe $t_0=4.32\times 10^{17}s$ ($\approx 13.7\times 10^9$ years). 
Also, in that equation, the time $t_h$ means the age of the Universe, when the accelerated expansion started, and $a_h$ gives the corresponding scale factor value, at that time.
We solve eq. (\ref{17}) for nine different set of values of \{$a_h$, $t_h$\}, finding nine different values for $\beta$. They are given in Table \ref{18}. After that, for each 
value of $\beta$ obtained, we solve eq. (\ref{3}), using the same values of the other parameters, as described in the previous paragraph. We use, as initial conditions to those equations, the 
appropriate values of $a_h$ and $t_h$. That gives the opportunity to obtain, for each value of $\beta$, the time it takes to the universe reaches the {\it big rip} singularity ($t_{br}$).
In other words to reach its end. The value of that time, for each value of $\beta$, is given in Table \ref{18}. Figure 7,
shows graphically the scale factor behavior as a function of $t$, for the nine NC models with different $\beta$ values.

\begin{figure}
\includegraphics[width=0.5\textwidth]{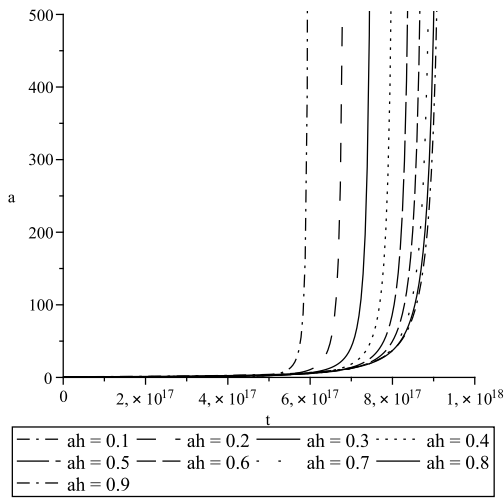}	
	\caption{Scale factor behavior as a function of $t$, for nine NC models with different values of $\beta$. The time $t$ is given in seconds.}
	\label{figure13}
	\end{figure}

\begin{table}[ht!]
	\centering
	\begin{tabular}{|c|c|c|c|} \hline
		$a_{h}$ & $t_{h} (Gyear)$ & $\beta$ & $t_{br} (Gyear)$ \\ \hline
		$0.9$ & $12.0224$ & $-1.280207183\times 10^{-35}$ & $29.9959$ \\ \hline
		$0.8$ & $10.5167$ & $-1.479734788\times 10^{-35}$ & $29.6796$ \\ \hline
		$0.7$ & $8.9511$ & $-1.764159846\times 10^{-35}$ & $29.1685$ \\ \hline
		$0.6$ & $7.3488$ & $-2.190432606\times 10^{-35}$ & $28.4103$ \\ \hline
		$0.5$ & $5.7470$ & $-2.870500485\times 10^{-35}$ & $27.3511$ \\ \hline
		$0.4$ & $4.1973$ & $-4.0483989954\times 10^{-35}$ & $25.9431$ \\ \hline
		$0.3$ & $2.7629$ & $-6.34762679\times 10^{-35}$ & $24.1457$ \\ \hline
		$0.2$ & $1.5148$ & $-11.839727650\times 10^{-35}$ & $21.9076$ \\ \hline
		$0.1$ & $0.5370$ & $-3.229384375\times 10^{-34}$ & $19.0753$ \\ \hline
	\end{tabular}
	\caption{{\protect\footnotesize Table with the estimates for $\beta$ and the time till the end of the Universe.}}
	\label{18}
	\end{table} 

From Table \ref{18}, we observe that $\beta$ increases as $t_h$ diminishes. That result agrees with the idea that noncommutativity must had been
more important at the beginning of the Universe. Another result, coming from Table \ref{18}, is that for smaller values of $\beta$, $t_{br}$
diminishes. It agrees with the conclusions of our study. For smaller values of $\beta$ the scale factor expands more rapidly. See Table \ref{6},
from the example of Subsection \ref{subsec1}, for another instance of that result.

Finally, it is important to mention that in Ref.\cite{gil3} the authors considered a NC FRW cosmological model, where the matter content is a
phantom perfect fluid. In Ref.\cite{gil3}, the NC was introduced by the following deformed Poisson brackets: $\{a_{nc}, P_{Tnc}\}=\{T_{nc}, P_{anc}\}=\gamma$,
where $\gamma$ is a NC parameter. Comparing those Poisson brackets with the deformed ones used here, eqs. (\ref{eq:19a}) and 
(\ref{eq:19a1}), we notice that they are different.
That difference between the two sets of Poisson brackets, leads to different scale factor dynamics for each NC theory. Therefore, they are not physically 
equivalent, as it was shown in a detailed comparison between those two theories\cite{gil2}. Although, we cannot directly compare our results with the ones in
Ref.\cite{gil3}, because the gauge we are considering here is different from the one used there, we notice that our estimates for the NC parameter are very 
different from the ones obtained there. That, in our opinion, may be another illustration of the difference between the two NC theories.

\section{Conclusions}
\label{conclusions}

Due to the presence of a phantom perfect fluid, in the model, and a negative NC parameter, the solutions to eq. (\ref{3}) will start at a 
minimum scale factor value, then, they will expand and finally reach an infinity scale factor value at a finite time. Those universes will end in a {\it big rip} singularity. 
It is important to mention that, the present noncommutativity does not remove the {\it big rip} singularity, already present in the corresponding commutative models. Solving 
eq. (\ref{3}) for many different values of all parameters, we reach the following conclusions concerning the scale factor behavior: (i) For any value of $\beta$, the scale factor in 
the NC model will always expand more rapidly than in the corresponding commutative one. Also, if we fix all parameters with the exception of $\beta$, the scale factor will expand
more rapidly in the NC model, for smaller values of that parameter; (ii) If we fix all parameters with the exception of $\alpha$, the scale factor will expand
more rapidly in the NC model, as well as, in the corresponding commutative one, for smaller values of that parameter; (iii) If we fix all parameters with the exception of $C$, the 
scale factor will expand more rapidly in the NC model, as well as, in the corresponding commutative one, for bigger values of that parameter; (iv) If we fix all parameters with the 
exception of $\Lambda$, the scale factor will expand more rapidly in the NC model, as well as, in the corresponding commutative one, for bigger values of that parameter; (v) If we 
fix all parameters with the exception of $k$, the scale factor will expand more rapidly in the NC model, as well as, in the corresponding commutative one, for $k=-1$, then for $k=0$
and finally for $k=1$; (vi) The dynamics of the scale factor also depends on the initial value of that quantity ($a(t=0)\equiv a_0$). We notice that, if we fix all parameters and let $a_0$ varies,
the scale factor will expand more rapidly in the NC model, as well as, in the corresponding commutative one, for bigger values of that quantity.

\section*{Acknowledgments}

L. G. Rezende Rodrigues thanks CAPES for his scholarship.


\begin{thebibliography} {99}

\bibitem{riess} A. G. Riess et al., Astron. J. {\bf 116}, 1009 (1998). 

\bibitem{perlmutter} S. Perlmutter et al., Astrophys. J. {\bf 517}, 565 (1999).

\bibitem{an} R. An, C. Feng and B. Wang, JCAP {\bf 10}, 49 (2017).

\bibitem{bilic} N. Bilic, G. B. Tupper and R. D. Viollier, Phys. Lett. B {\bf 535}, 17-21 (2002).

\bibitem{bertolami} M. C. Bento, O. Bertolami and A. A. Sen, Phys. Rev. D {\bf 66}, 043507 (2002).

\bibitem{caldwell} R. R. Caldwell, Phys. Lett. B {\bf 545}, 23-29 (2002).

\bibitem{peebles} P. J. E. Peebles and B. Ratra, Rev. Mod. Phys. {\bf 75}, 559 (2003).

\bibitem{caldwell1} R. R. Caldwell, M. Kamionkowski and N. N. Weinberg, Phys. Rev. Lett. {\bf 91}, 071301 (2003).

\bibitem{copeland} E. J. Copeland, M. Sami and S. Tsujikawa, Int. J. Mod. Phys. D {\bf 15}, 1753–1935 (2006).

\bibitem{padmanabhan} T. Padmanabhan, Gen. Relativ. Gravit. {\bf 40}, 529–564 (2008).

\bibitem{cai} Y. F. Cai, E. N. Saridakis, M. R. Setare, J. Q. Xia, Phys. Rep. {\bf 493}, 1-60 (2010).

\bibitem{Mli} M. Li, X. D. Li, S. Wang and S. Wang, Commun. Theor. Phys. {\bf 56}, 525–604 (2011).

\bibitem{snyder} H. S. Snyder, Phys. Rev. {\bf 71}, 38 (1947).

\bibitem{douglas} M. R. Douglas and N. A. Nekrasov, Rev. Mod. Phys. {\bf 73}, 977 (2001).

\bibitem{schwarz} A. Konechny and A. Schwarz, Phys. Rep. {\bf 360}, 353-465 (2002). 

\bibitem{szabo} R. J. Szabo, Phys. Rep. {\bf 378}, 207-299 (2003).

\bibitem{banerjee} R. Banerjee, B. Chakraborty, S. Ghosh, P. Mukherjee, S. Samanta, Found. Phys. {\bf 39}, 1297 (2009).

\bibitem{garcia} H. Garcia-Compean, O. Obregon and C. Ramirez, Phys. Rev. Lett. {\bf 88}, 161301 (2002).

\bibitem{nelson} G. D. Barbosa and N. Pinto-Neto, Phys. Rev. D {\bf 70}, 103512(2004).

\bibitem{barbosa} G. D. Barbosa, Phys. Rev. D {\bf 71}, 063511 (2005).

\bibitem{sabido} W. Guzm\'{a}n, M. Sabido and J. Socorro, Phys. Rev. D {\bf 76}, 087302 (2007).

\bibitem{gil} G. Oliveira-Neto, M. Silva de Oliveira, G. A. Monerat and E. V. Corr\^{e}a Silva, Int. J. Mod. Phys. D {\bf 26}, 1750011 (2016).

\bibitem{pedram} B. Vakili, P. Pedram and S. Jalalzadeh, Phys. Lett. B {\bf 687}, 119 (2010).

\bibitem{obregon} O. Obregon and I. Quiros, Phys. Rev. D {\bf 84}, 044005 (2011).

\bibitem{sabido1} W. Guzm\'{a}n, M. Sabido and J. Socorro, Phys. Lett. B {\bf 697}, 271 (2011).

\bibitem{sabido2} S. P\'{e}rez-Pay\'{a}n, M. Sabido and C. Yee-Romero, Phys. Rev. D {\bf 88}, 027503 (2013).

\bibitem{gil2} G. A. Monerat, E. V. Corr\^{e}a Silva, C. Neves, G. Oliveira-Neto, 
L. G. Rezende Rodrigues and M. Silva de Oliveira, Int. J. Mod. Phys. D {\bf 26}, 1750022 (2016).

\bibitem{gil3} G. Oliveira-Neto and A. R. Vaz, Eur. Phys. J. Plus {\bf 132}, 131 (2017).

\bibitem{gil0} E.V. Corr\^{e}a Silva, G. A. Monerat G. Oliveira-Neto C. Neves and L. G. Ferreira Filho, Phys. Rev. D {\bf 80}, 047302 (2009).

\bibitem{gil1} E. M. C. Abreu, M. V. Marcial, A. C. R. Mendes, W. Oliveira and G. Oliveira-Neto,
JHEP {\bf 05}, 144 (2012).

\bibitem{gil4} E. M. C. Abreu, A. C. R. Mendes, G. Oliveira-Neto, J. Ananias Neto, L. G. Rezende Rodrigues and M. Silva de Oliveira, 
Gen. Relativ. Gravit. {\bf 51}, 95 (2019).

\bibitem{sabido3} J. C. L\'{o}pez-Dom\'{\i}nguez, O. Obreg\'{o}n, M. Sabido and C. Ram\'{\i}rez, Phys. Rev. D {\bf 74}, 084024 (2006).

\bibitem{sabido4} L. F. Escamilla-Herrera, E. A. Mena-Barboza and J. Torres-Arenas, Entropy {\bf 18}, 406 (2016).

\bibitem{sabido5} A. Crespo-Hern\'{a}ndez, E. A. Mena-Barboza and M. Sabido, Entropy {\bf 19}, 91 (2017).

\bibitem{faddeev} L. Faddeev and R. Jackiw, {\it Hamiltonian reduction of unconstrained and constrained systems},
Phys. Rev. Lett. 60 (1988) 169.

\bibitem{barcelos} J. Barcelos-Neto and C. Wotzasek, {\it Symplectic quantization of constrained systems},
Mod. Phys. Lett. A 7 (1992) 1737. 

\bibitem{barcelos1} J. Barcelos-Neto and C. Wotzasek, {\it Faddeev-Jackiw quantization and constraints},
Int. J. Mod. Phys. A 7 (1992) 49.

\bibitem{montani} H. Montani, C. Wotzasek, Mod. Phys. Lett. A 8, 3387 (1993).

\bibitem{misner} R. Arnowitt, S. Deser and C. W. Misner, in {\it Gravitation: an introduction to current 
research}, ed. L. Witten (Wiley, New York, 1962), Chapter 7, pp 227-264 and arXiv:gr-qc/0405109.

\bibitem{schutz} Schutz, B. F., Phys. Rev. D \textbf{2}, 2762, (1970).

\bibitem{schutz1} Schutz, B. F., Phys. Rev. D \textbf{4}, 3559, (1971).

\bibitem{rubakov} V. G. Lapchinskii and V. A. Rubakov, Theor. Math. Phys. {\bf 33}, 1076 (1977).

\bibitem{germano1} F. G. Alvarenga, J. C. Fabris, N. A. Lemos, G. A. Monerat, Gen. Rel. Grav. {\bf 34}, 651 (2002).

\bibitem{gil5} G. A. Monerat, G. Oliveira-Neto, E.V. Corr\^{e}a Silva, L. G. Ferreira Filho, P. Romildo, Jr., J. C. Fabris,
R. Fracalossi, S. V. B. Gonc\c alves and F. G. Alvarenga, Phys. Rev. D {\bf 76}, 024017 (2007).

\bibitem{riess1} A. G. Riess, L. Macri, S. Casertano, H. Lampeitl, H. C. Ferguson, A. V. Filippenko,
S. W. Jha, W. Li, J. M. Silverman and R. Chornock, Ap. J. {\bf 730}, 119 (2011).

\end{thebibliography}
\end{document}